\documentstyle[aps,twocolumn,epsf,tighten]{revtex}
\begin{document}
\draft
\title
{\hfill\begin{minipage}{0pt}\scriptsize \begin{tabbing}
\hspace*{\fill} HU-EP-00/17\\
\hspace*{\fill} WUB 00-06\\ 
\hspace*{\fill} HLRZ 2000-4\end{tabbing} 
\end{minipage}\\[8pt]  
Static potentials and glueball masses from QCD simulations
with Wilson sea quarks}
\author{Gunnar S.\ Bali\thanks{Electronic mail: bali@physik.hu-berlin.de},}
\address{Institut f\"ur Physik, Humboldt-Universit\"at zu Berlin,
Invalidenstr.\ 110, D-10115 Berlin, Germany and\\
Department of Physics \& Astronomy, The University of Glasgow, Glasgow G12 8QQ,
Scotland}
\author{Bram Bolder, Norbert Eicker,
Thomas Lippert, Boris Orth, Peer Ueberholz,}
\address{Fachbereich Physik, Bergische Universit\"at
Wuppertal, D-42097 Wuppertal, Germany}
\author{Klaus Schilling and Thorsten Struckmann}
\address{NIC, Forschungszentrum J\"ulich, D-52425 J\"ulich and\\
DESY, D-22603 Hamburg, Germany}
\author{(SESAM and T$\chi$L Collaborations)}

\date{\today}
\maketitle

\begin{abstract}
We calculate glueball and torelon masses as well as
the lowest lying hybrid potential in addition
to the static ground state potential
in lattice simulations of QCD with two flavours of dynamical
Wilson fermions. The results are obtained on lattices
with $16^3\times 32$ and $24^3\times 40$ sites at $\beta=5.6$,
corresponding to a lattice spacing, $a^{-1}=2.65^{+5}_{-8}$~GeV,
as determined from the Sommer force radius,
at physical sea quark mass.
The range spanned in the present study of
five different quark masses
is reflected in the ratios,
$0.83\geq m_{\pi}/m_{\rho}\geq 0.57$.


\end{abstract}

\narrowtext
\section{Introduction}
The expectation that gluons form bound states, so-called
glueballs, is as old as QCD itself~\cite{Fritzsch:1973pi}. Indeed,
such states make up the spectrum of pure $SU(3)$ gauge theory
and have rather precisely been determined in
quenched lattice
simulations~\cite{Bali:1993fb,Sexton:1995kd,Morningstar:1999rf}.
However, an unambiguous experimental confirmation of their existence
is still missing. Lattice simulations have provided an explanation
for this situation: most of the glueballs
come out to be quite heavy~\cite{Bali:1993fb,Morningstar:1999rf}
and will therefore manifest themselves as very broad resonances with
many decay channels.
This holds in particular
for the most interesting glueballs, those
with spin-exotic, i.e.\ quark model forbidden,
quantum numbers.

However, a conservative interpretation
of results obtained in the valence quark (or quenched) approximation
to QCD
urges us to expect (see e.g.\ Ref.~\cite{Bali:2000gf}) 
a predominantly gluonic bound state with
scalar quantum numbers, $J^{PC}=0^{++}$,
and mass between 1.4~GeV and 1.8~GeV. Indeed, in this region
more scalar resonances have been established experimentally
than a standard quark model classification would
suggest~\cite{Close:1997nj}.
When switching on light quark flavours the difference between a
glueball that contains sea quarks and a
flavour singlet (isoscalar) meson that contains ``glue'',
sharing the same quantum numbers, becomes ill defined.
In general such hypothetical, pure states will mix with each other
to yield the observed hadron spectrum.
This situation is different from that of flavour
non-singlet hadrons which, if we ignore weak interactions,
have a well defined valence
quark content.
Such ``standard'' hadrons might
become unstable but otherwise
retain most of their qualitative properties
when sea quarks are included. Indeed ratios
between masses of light standard
hadrons, calculated within the quenched approximation,
have been found to differ by less than about~\cite{Aoki:2000yr} 10~\% 
from (full QCD) experiment.

One can utilise quenched lattice data on the
scalar glueball and isoscalar $s\bar{s}$ as well
as $u\bar{u}+d\bar{d}$ mesons
as an input
into phenomenologically motivated mixing models~\cite{Lee:2000kv}.
However, such models can only provide a crude scenario
for the
actual situation that will be clarified
once the entire spectrum of full QCD has been calculated.
A first attempt in this direction has been made by
the HEMCGC Collaboration a decade ago~\cite{Bitar:1991wr}
and recently first results 
from the UKQCD Collaboration on
flavour singlet scalar meson/glueball masses
in QCD with two light sea quark flavours have been
reported~\cite{Michael:1999rs}.

At present the main target of simulations involving sea quarks
is to pin down differences with respect to the quenched approximation
and to develop the methodology required when a decrease
of sea quark masses below $m_{\pi}/m_{\rho}\approx 0.5$ will become
possible with the advent of
the next generation of supercomputers.
One would expect sea quarks
not only to be important for glueball-meson mixing
and for an understanding of the spectrum of isoscalars but
for
flavour singlet phenomenology in general~\cite{Gusken:1999te,Gusken:1999as}.

The potential between static colour sources at a separation, $r$,
is one of the most precisely determined quantities in quenched lattice
studies~\cite{Bali:1992ab,Bali:1993ru,Booth:1992bm}.
Two physical phenomena are expected to occur when including sea quarks,
one at large and the other at small distances.
The former effect is known as ``string breaking''
(see Ref.~\cite{Schilling:1999mv} and references therein):
once $r$ exceeds
a critical value, creation of a light quark-antiquark pair
out of the vacuum becomes energetically preferable and the QCD
``string'' will break; the static-static state will decay into
a pair of static-light mesonic bound states, resulting in
the complete screening of colour charges at large distances that is observed
in nature. The potential will approach a constant value at infinite $r$.
On the other hand, the presence of sea quarks slows down the
``running'' of the QCD coupling as a function of the scale
with respect to the quenched approximation: when running the coupling
from an infra-red hadronic reference scale down to short distances, the
effective Coulomb coupling in presence of sea quarks should, therefore,
remain stronger than in the quenched case.

Exploratory studies~\cite{Born:1994cq,Heller:1994rz}
of the zero temperature static potential with dynamical fermions
focussed on the question of string breaking. However, no statistically
significant signal of colour screening has been detected.
A difference between the potentials with
and without sea quarks at short distances
has then been
reported by the SESAM Collaboration
in Refs.~\cite{Glassner:1996xi,Bali:1998bj,Gusken:1998sa}.
Later studies by the UKQCD, CP-PACS and MILC collaborations
qualitatively
confirm these findings~\cite{Garden:1999hs,Aoki:1999sb,Bernard:2000gd},
employing different lattice actions and, in the latter case, a
different number of flavours.

In this paper we present a combined analysis of SESAM and
T$\chi$L~\cite{Lippert:1998vg}
data (as obtained on $16^3 \times 32$ and $24^3 \times 40$
lattices, respectively) on the static potential, torelon and
glueball masses. Moreover, for the first time, we have determined the $\Pi_u$
hybrid potential~\cite{Griffiths:1983ah} in a simulation including dynamical
sea quarks. Preliminary versions of the results presented here have been
published in
Refs.~\cite{Glassner:1996xi,Bali:1997ec,Bali:1998de,Bali:1998bj,Gusken:1998sa,Schilling:1999mv,Bali:1999gq,Bali:2000gf}.

The article is organised as follows: in Sect.~\ref{sec:lat}
we summarise simulation details and our parameter values and describe
the measurement techniques applied. In Sect.~\ref{sec:mass} we discuss
how the glueball and torelon masses and the potentials are determined.
We also present the quality of the ground state overlaps achieved and
the forms of the respective creation operators that were found to be
most suitable.
Subsequently, we present our physical results on
the potentials, torelons and glueballs
in Sect.~\ref{sec:results}, before we conclude.

\section{Lattice technology}
\label{sec:lat}
We analyse samples of configurations that have been generated
by means of the hybrid Monte Carlo (HMC) algorithm
using the Wilson fermionic and gluonic actions with $n_f=2$
mass degenerate quark flavours at the inverse lattice coupling,
$\beta=5.6$. This is done on
$L_{\sigma}^3L_{\tau}=16^332$ as well as on $24^340$ lattices at the
mass parameter values,
$\kappa = 0.156, 0.1565, 0.157, 0.1575$ and 0.158.
The corresponding
chiralities can be quantified in terms of
the ratios~\cite{inprep2},
$m_{\pi}/m_{\rho}=0.834(3), 0.813(9), 0.763(6), 0.704(5)$ and
0.574(13), respectively.
The
simulation parameters are displayed in Tab.~\ref{tab:simul}, where the Sommer
scale~\cite{Sommer:1994ce}, $r_0\approx 0.5$ fm, is defined through the
static potential, $V(r)$,
\begin{equation}
\label{eq:som}
r^2\left.\frac{dV}{dr}\right|_{r=r_0}=1.65.
\end{equation}

At each $\kappa$ value 4,000--5,000 thermalised HMC trajectories have
been generated.
Great care has been spent~\cite{Alles:1998jq,Lippert:1999eg}
on investigating autocorrelations
between successive HMC trajectories for the topological charge
as well as for other quantities including
smeared Wilson loops and Polyakov lines, similar to those
used in the present study.
The operators that enter  the glueball and torelon analysis have been
separated by two (eight at $\kappa=0.1565$, five at $\kappa=0.157$)
HMC trajectories;
smeared Wilson loops that are required for the potential
calculations
have been determined every 20 (16 at $\kappa=0.1565$)
trajectories. The numbers of thermalised configurations
analysed in
total are denoted by
$n_{\mbox{\scriptsize glue}}$ and
$n_{\mbox{\scriptsize pot}}$ in the Table for the two classes of
measurements, respectively.

In order to reduce statistical noise we had to employ
measurement frequencies that are
larger than the inverse integrated
autocorrelation times in most cases. This is particularly
true for the glueballs and torelons. 
Autocorrelation effects have been taken care of by
binning the time series
into blocks
prior to the statistical analysis. The bin sizes have been
increased until the statistical errors of the fitted parameters
stabilised.
The numbers
of effectively independent configurations that,
after this averaging process, finally
entered the analysis
are denoted by $m_{\mbox{\scriptsize glue}}$ and
$m_{\mbox{\scriptsize pot}}$, respectively.

In addition to the dynamical quark simulations, quenched reference
potential measurements have been performed. For this purpose,
smeared Wilson loop data generated at $\beta=6.0$ and
$\beta=6.2$ in the context of the study of
Ref.~\cite{Bali:1997am} have been re-analysed.
The hybrid $\Pi_u$ potential reference data have been obtained
in Ref.~\cite{Collins:1998cb} at the same two $\beta$ values. 
\begin{table}
\caption{Simulation parameters. The last two rows refer to
quenched simulations.}
\label{tab:simul}
\begin{tabular}{cccccccc}
$\kappa$&$L_{\sigma}^3L_{\tau}$&$r_0a^{-1}$&$m_{\pi}aL_{\sigma}$&
$n_{\mbox{\scriptsize glue}}$&$m_{\mbox{\scriptsize glue}}$&
$n_{\mbox{\scriptsize pot}}$&$m_{\mbox{\scriptsize pot}}$\\
\hline
0.1560&$16^332$&5.11(3)&7.14(4)&2128&266&236&236\\
0.1565&$16^332$&5.28(5)&6.39(6)&500&250&322&161\\
0.1570&$16^332$&5.48(7)&5.51(4)&1000&200&250&125\\
0.1575&$16^332$&5.96(8)&4.50(5)&2272&142&270&90\\
0.1575&$24^340$&5.89(3)&6.65(6)&1980&110&201&67\\
0.1580&$24^340$&6.23(6)&4.77(7)&1780&89&196&49\\\hline
$\beta=6.0$&$16^332$&5.33(3)& --- & --- & --- &570&570\\
$\beta=6.2$&$32^4$&7.29(4)& --- & --- & --- &116&116\\
\end{tabular}
\end{table}

In order to determine glueball masses, temporal
correlation functions between linear combinations of Wilson loops
have been constructed. For simplicity we restricted
ourselves to plaquette-like
operators that can be built from four (fat) links.
Torelons, i.e.\ flux loops that live on the torus and encircle a periodic
spatial boundary~\cite{Michael:1989vh}, have been created from
linear combinations of
smeared spatial Polyakov (or Wilson) lines.
Both, the fat plaquettes, used to construct the glueballs,
and the Wilson lines can have three orthogonal spatial orientations.
The real part of the sum over these
orientations transforms according to the
$A_1^{++}$ representation~\cite{Berg:1983kp}
of the relevant cubic point group,
$O_h\otimes Z_2$. In order to project out the $T_1^{+-}$ representation
one can take the imaginary part of one of the three possible orientations;
the $E^{++}$ representation can be achieved by taking the real
part, either of the sum of two
orientations minus twice the third one or
of the difference between two orientations.
In order to increase statistics we average over the correlation functions
obtained from the three equivalent
orthogonal $T_1^{+-}$ or two $E^{++}$ realisations.
Within all three channels, $A_1^{++}$, $E^{++}$ and $T_1^{+-}$,
we subtract the disconnected parts
from the correlation functions. This is only necessary
for the $A_1^{++}$ state that carries the vacuum quantum numbers.
However, in doing this for all channels, we find (slightly)
reduced statistical fluctuations.

All operators are projected onto zero momentum at source and sink
by averaging over all spatial sites.
In the case of glueballs, the $A_1^{++}$, $E^{++}$ and $T_1^{+-}$
representations
can be subduced from the continuum $O(3)\otimes Z_2$,
$J^{PC}=0^{++},2^{++}$ and $1^{+-}$
representations, respectively, while
the cubic point group remains the relevant symmetry group
for the torelons (that only exist on a torus), even in the continuum limit.
In what follows we shall label glueballs by the above continuum
$J^{PC}$ quantum numbers.

The fat links that form the basis of the glueball and torelon
creation operators have been
constructed by alternating ``APE smearing''~\cite{Albanese:1987ds}
(keeping the length of the smeared link constant) with
``Teper fuzzing''~\cite{Teper:1987wt} (increasing the length by a factor
two). In both algorithms a link of smearing level $i+1$ is
formed by taking 
a linear combination of its predecessor of iteration $i$ (in the 
case of fuzzing
the product of two straight $i$ level links) and the four closest
surrounding spatial staples, followed by a projection
back into the gauge group~\cite{Bali:1992ab}. We denote the un-smeared
links by $i=0$. The first iteration then consists of APE smearing
with the relative weight of the straight line connection
with respect to the staple sum
being $\alpha=0$. This is followed by a fuzzing iteration
with $\alpha=1.3$, APE smearing with $\alpha=0$ and so forth.
This procedure is the outcome of a semi-systematic
study of different
fuzzing algorithms~\cite{Bali:1997ec}.
Fat links of levels 0 and 1 have effective length $a$,  levels
2 and 3 length $2a$,  levels 4 and 5 length $4a$ and levels 6 and 7
length $8a$. Iterations 8 and 9, yielding fat
links of extent $16a$,
are only performed on the $24^3$ lattices.

APE smearing has also been applied
to obtain ground state and excited state potentials.
In this case, we iterate
the procedure 26 times with a straight line coefficient,
$\alpha=2.3$. Subsequently, Wilson loops are constructed
whose spatial parts are built from smeared links.
In addition to on-axis distances, ${\mathbf r}\parallel (1,0,0)$,
off-axis separations that are
multiples of $(1,1,0)$, $(1,1,1)$, $(2,1,0)$,
$(2,1,1)$ and $(2,2,1)$ have been realised for the ground state potential.
In the case of the hybrid potential, in addition to
on-axis separations, the plane-diagonal direction, ${\mathbf r}\parallel
(1,1,0)$,
has been investigated. The spatial gauge transporter along the latter
direction is obtained by
calculating the difference between the positively and negatively oriented
``one-corner'' paths connecting the two end points.
For on-axis separations the potential in the continuum
$\Pi_u$ representation can be obtained from the
$E_u$ representation of $D_{4h}$. The $E_u$ creation operator
has been
constructed in the usual way~\cite{Campbell:1984fe}
from the sum of two differences between
forwards and backwards oriented staples. The orthogonal depth
of these staples has been chosen to be one lattice unit for $r=a$,
two lattice units for $2a\leq r\leq 5a$ and three lattice units for
$r\geq 6a$.

\section{Mass determinations}
\label{sec:mass}
A cross correlation matrix between the eight (ten in the case of
the $L_{\sigma}=24$ glueballs) basis operators corresponding to
different smearing/fuzzing levels is formed for each temporal separation,
$t$. For the sake of numerical
stability we restrict our analysis to two by two subsets\footnote{
In the case of the
$1^{+-}$ glueball only the diagonal elements of the correlation matrix
are used.}
of the correlation matrix, $M(t)$, with blocking level $i\geq 2$.
We then select the subset of two states for which the combination,
$M^{-1/2}(a)M(2a)M^{-1/2}(a)$, yields the lowest energy
eigenvalue, and
trace the $t$ dependence of the
effective mass,
\begin{equation}
m_{\mbox{\scriptsize eff}}(t)=a^{-1}\ln\frac{C(t)}{C(t+a)},
\end{equation}
calculated from the correlation function, $C(t)$,
that is associated with the optimal
eigenvector, in order to identify a plateau.
The glueball and torelon masses are
subsequently determined by means of a fully correlated and
bootstrapped fit to the plateau region. The required covariance matrices
are determined by means of sub-bootstraps.

We denote the state that is created by
an operator of fuzzing level $i=0,1,\ldots$ within each glueball
channel by $|i\rangle$.
In Tabs.~\ref{tab:wave} and \ref{tab:wave2} the
state vectors used to extract the
$0^{++}$ and $2^{++}$ glueballs are listed, respectively,
as well as the fitted overlaps of these
states with the physical ground states.
We have employed the normalisation, $\langle i|i\rangle=1$.
The data at $\kappa=0.1565$ is less precise due to the smaller
ensemble size, $n_{\mbox{\scriptsize glue}}$.
In general, our variational procedure
yields ground state overlaps around 80~\%.

While a direct determination
of glueball radii turns out to be impossible within
our limited statistical ensemble sizes,
the shapes of the optimal creation operators are indicative for
the approximate diameters of the underlying
physical ground states. 
Note that the only direct measurement of
glueball sizes to-date has been
attempted in pure $SU(2)$ gauge theoryin Ref.~\cite{Forcrand:1992kc}.
On all $16^3$ volumes we find the
$0^{++}$ wave function to receive a dominant $|5\rangle$ component,
which is one APE iteration applied to a fuzzed link of effective length $4a$.
The subleading contribution is $|6\rangle$, i.e.\ a square with side length
$8a$.
We conclude that the ``diameter'' of the scalar glueball
is somewhat less than 8 lattice spacings or 0.7~fm.
These observations indicate a somewhat wider glueball than
observed in previous
quenched
studies~\cite{Michael:1989jr,Gupta:1991ek,Forcrand:1992kc,Bali:1993fb}.

As one increases $\kappa$,
the effective
lattice spacing, determined for instance in units of the gluonic
observable, $r_0$, decreases and
the glueball extends over more lattice sites.
This results in an increased relative
weight of the larger, $|6\rangle$, state, which finally becomes
the dominant contribution at $\kappa=0.158$. There is a difference
between the two volumes at $\kappa=0.1575$: the size of
the glueball
on the larger lattice becomes smaller, indicating
finite size effects. As we shall see below this goes along
with an increase of its mass.

The $2^{++}$ state is dominated by the $|6\rangle$ contribution, with
the exception of $\kappa=0.1565$. At this $\kappa$ value the
statistics are not really sufficient for a reliable determination of the
optimal state vector, as indicated by the large relative error
on the ground state overlap, $0.48\pm 0.20$. 
The second largest contribution then comes from the $|4\rangle$
($|3\rangle$ for the $\kappa=0.1575$, $L_{\sigma}=24$ simulation)
state. In general, we find the $|4\rangle$ state to have best overlap with
the $1^{+-}$ glueball.

\begin{table}
\caption{Scalar glueball wave function and overlap with the physical ground state.}
\label{tab:wave}
\begin{tabular}{cccc}
$\kappa$&$L_{\sigma}$&state vector&g.s.\ overlap\\\hline
0.1560&$16$&$0.76|5\rangle+0.65|6\rangle$&0.89(2)\\
0.1565&$16$&$0.77|5\rangle+0.63|6\rangle$&0.71(12)\\
0.1570&$16$&$0.72|5\rangle+0.70|6\rangle$&0.79(4)\\
0.1575&$16$&$0.73|5\rangle+0.68|6\rangle$&0.83(5)\\
0.1575&$24$&$0.72|4\rangle+0.70|5\rangle$&0.94(9)\\
0.1580&$24$&$0.65|5\rangle+0.76|6\rangle$&0.80(8)
\end{tabular}
\end{table}

\begin{figure}
\centerline{\epsfxsize=8cm\epsfbox{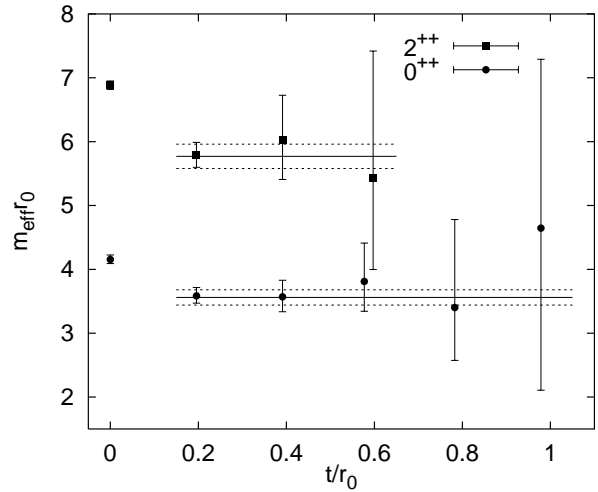}}
\caption{Effective scalar and tensor glueball masses and fits at
$\kappa=0.156$.}
\label{fig:glue1}
\end{figure}

\begin{figure}
\centerline{\epsfxsize=8cm\epsfbox{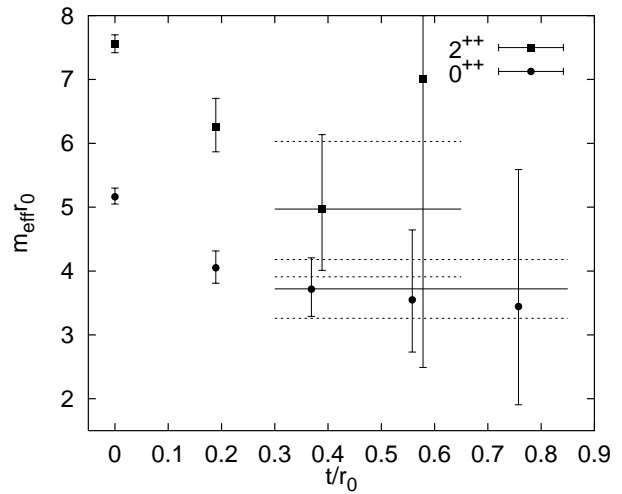}}
\caption{Same as Fig.~\protect\ref{fig:glue1} for $\kappa=0.1565$.}
\label{fig:glue2}
\end{figure}

\begin{table}
\caption{Tensor glueball wave function and overlap with the physical ground state.}
\label{tab:wave2}
\begin{tabular}{cccc}
$\kappa$&$L_{\sigma}$&state vector&g.s.\ overlap\\\hline
0.1560&$16$&
$0.37|4\rangle+0.93|6\rangle$&0.81(3)\\
0.1565&$16$&
$0.83|4\rangle+0.55|6\rangle$&0.48(20)\\
0.1570&$16$&
$0.28|4\rangle+0.96|6\rangle$&0.81(3)\\
0.1575&$16$&
$0.27|4\rangle+0.96|6\rangle$&0.80(3)\\
0.1575&$24$&
$0.37|3\rangle+0.93|6\rangle$&0.62(17)\\
0.1580&$24$&
$0.66|4\rangle+0.75|6\rangle$&0.84(4)
\end{tabular}
\end{table}

\begin{figure}
\centerline{\epsfxsize=8cm\epsfbox{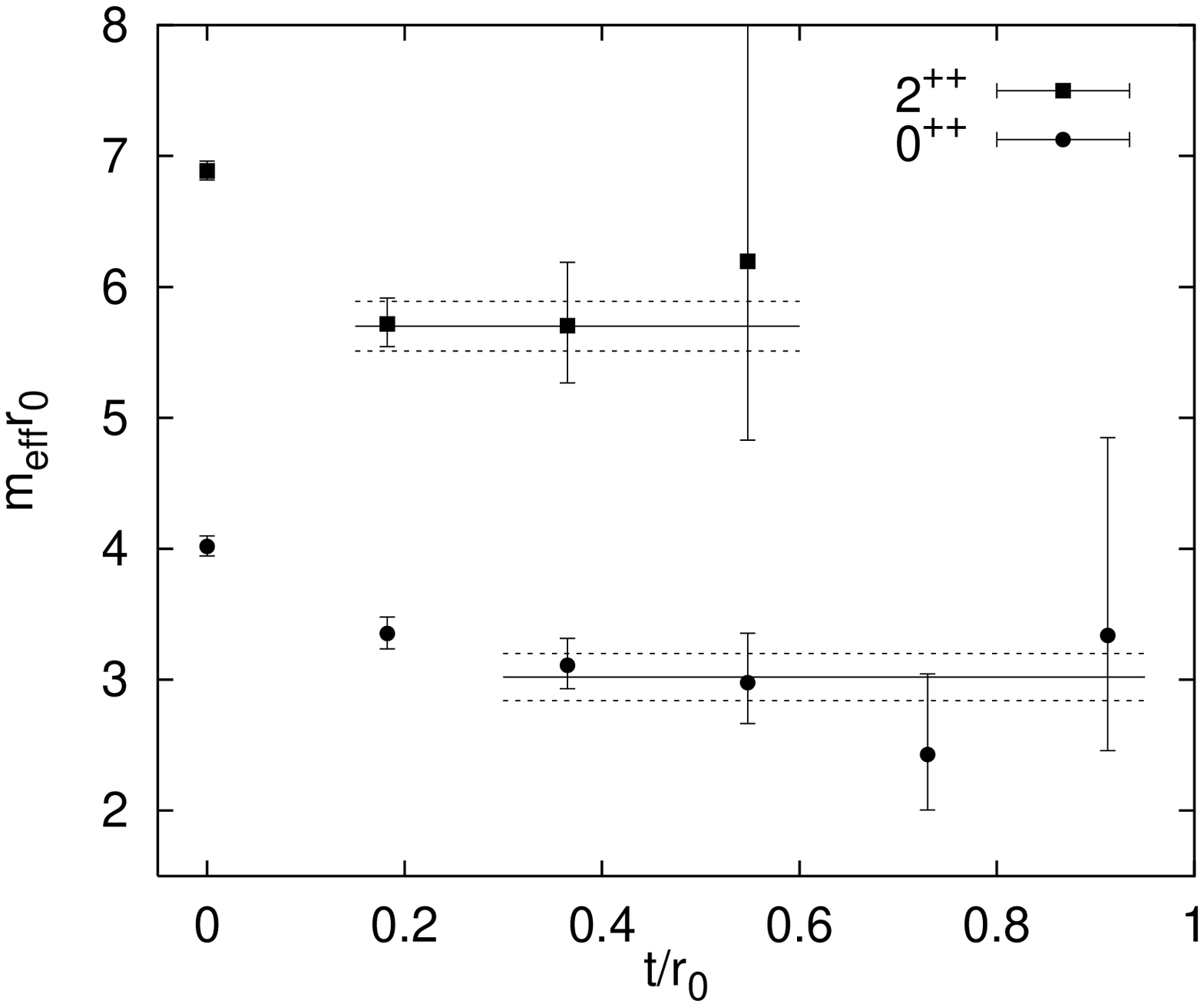}}
\caption{Same as Fig.~\protect\ref{fig:glue1} for $\kappa=0.157$.}
\label{fig:glue3}
\end{figure}

\begin{figure}
\centerline{\epsfxsize=8cm\epsfbox{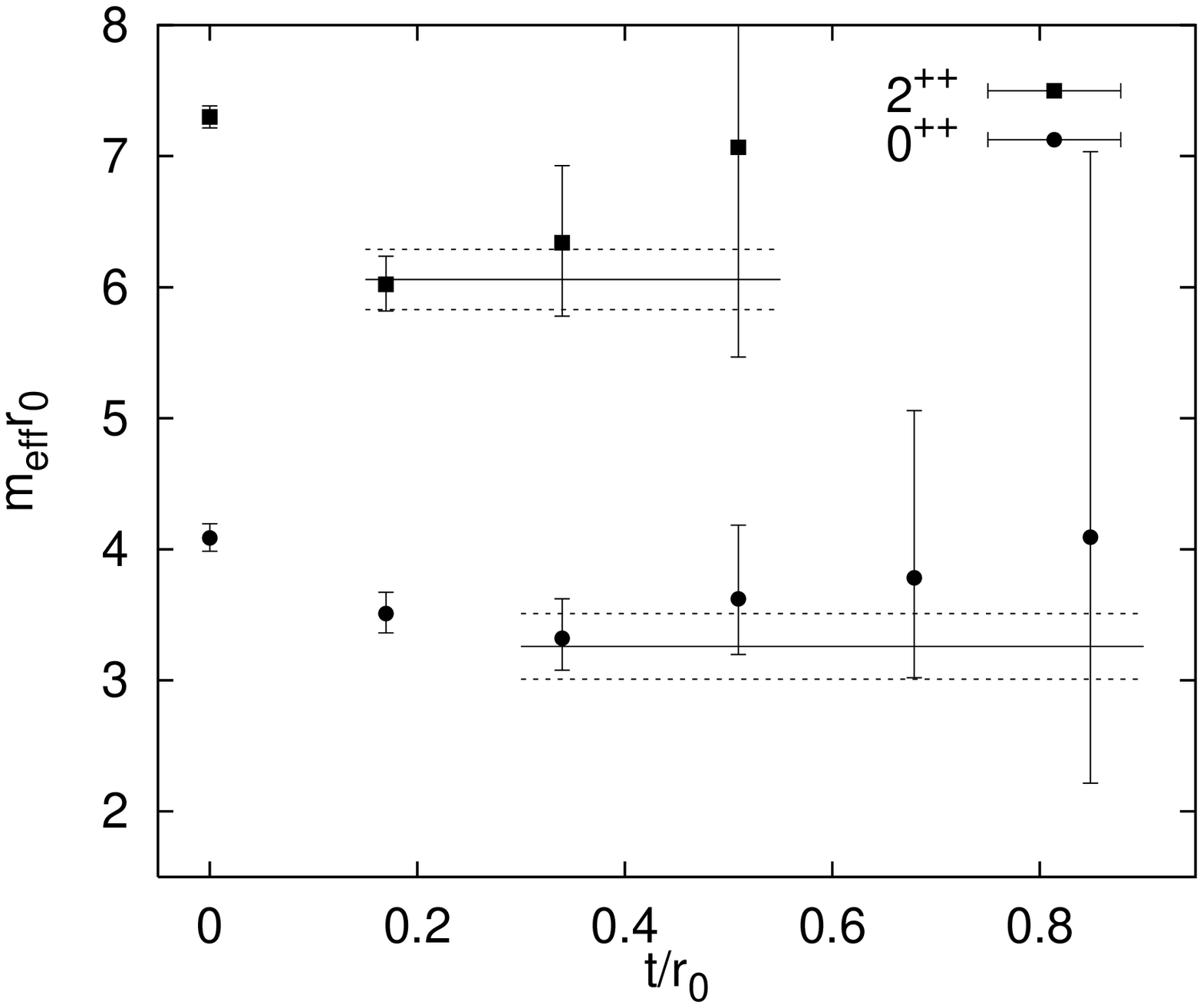}}
\caption{Same as Fig.~\protect\ref{fig:glue1} for the $16^3$ volume
at $\kappa=0.1575$.}
\label{fig:glue4}
\end{figure}

\begin{table}
\caption{Torelon ground state overlaps.}
\label{tab:over2}
\begin{tabular}{ccccc}
$\kappa$&$L_{\sigma}$&$A_1^{++}$&$E^{++}$&$T_1^{+-}$\\\hline
0.1560&16&0.89(2)&0.90(2)&0.91(2)\\
0.1570&16&0.87(6)&0.82(4)&0.72(11)\\
0.1575&16&0.41(10)&0.74(2)&0.39(12)\\
0.1575&24&0.55(13)&0.84(3)&0.60(21)\\
0.1580&24&0.85(24)&0.83(11)&0.24(9)
\end{tabular}
\end{table}

\begin{figure}
\centerline{\epsfxsize=8cm\epsfbox{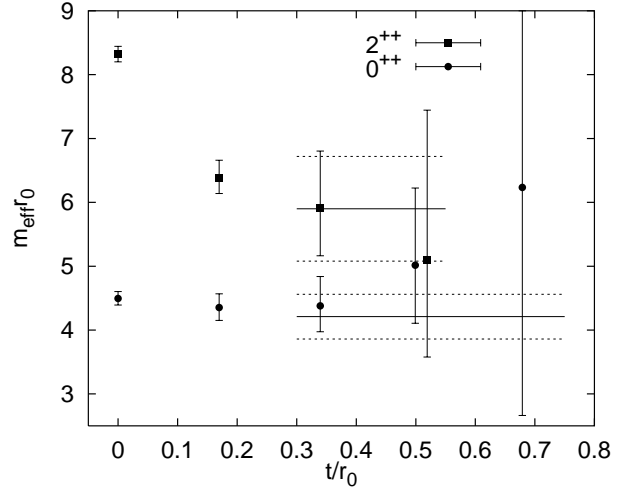}}
\caption{Same as Fig.~\protect\ref{fig:glue1} for the $24^3$ volume
at $\kappa=0.1575$.}
\label{fig:glue5}
\end{figure}

\begin{figure}
\centerline{\epsfxsize=8cm\epsfbox{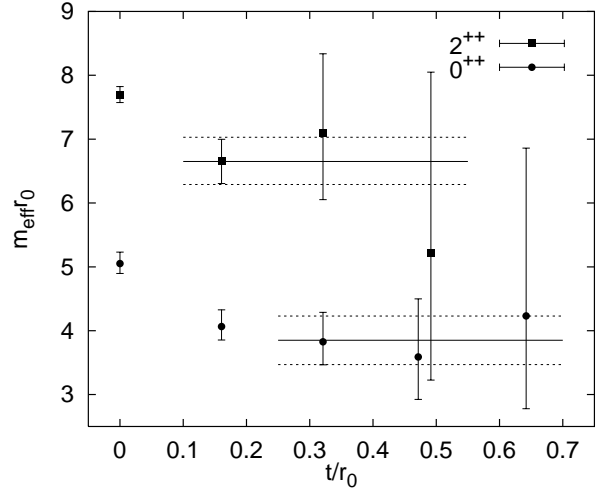}}
\caption{Same as Fig.~\protect\ref{fig:glue1} for $\kappa=0.158$.}
\label{fig:glue6}
\end{figure}

In Figs.~\ref{fig:glue1} -- \ref{fig:glue6}, we show effective mass plots
for the scalar and tensor glueballs at the various parameter values,
together with the corresponding fit results (solid lines with
dashed error bands).
Plateaus could be identified either from $t=2a$ or $t=3a$ onwards.
Since we take correlations between the data points into account
fitted results can
turn out to lie somewhat below the central values of the
individual effective masses (cf.\ Figs.~\ref{fig:glue4} and \ref{fig:glue5}).
The torelon masses have been analysed in an analogous way. All torelons
receive their
dominant contribution either from the $|6\rangle$ or $|7\rangle$ channels
with a subleading term from the $|4\rangle$ or $|5\rangle$ states.
The achieved ground state overlaps are listed in Tab.~\ref{tab:over2}.
The statistical quality of the signal did not permit an estimation of
the overlap at $\kappa=0.1565$.

On volumes of similar physical size in terms of $r_0$ we find
statistical errors of fuzzed plaquettes and Wilson lines
to be smaller than in comparable quenched
simulations~\cite{Michael:1989jr,Bali:1993fb}.
This can very well be due to the larger number of degrees of freedom
when including quarks. Similar conclusions can be drawn from a comparison
between published $SU(2)$ and $SU(3)$ results. In the case of the
determination of the static potential absolute errors on smeared Wilson
loops are smaller than in comparable quenched simulations too. However,
this effect is partly compensated for by a faster decay of the signal in
Euclidean time, due to an increase of the static source self energy.
In most gluonic quantities self-averaging over
more lattice points when increasing the physical volume reduces
the statistical noise. This is of course not the case for torelons that
become heavier with growing volume. However,
statistical fluctuations of glueball correlation functions
grow with the volume too.
This effect, that has also been observed in quenched
simulations~\cite{Michael:1989jr,Bali:1993fb}, is related to contributions
stemming
from large distance correlators to the spatial sum that is required to project
onto zero momentum.

In the potential measurements no variational optimisation is performed.
Again,
effective masses are traced against $t$, separately for all
lattice separations, ${\mathbf r}$.
In view of the fact
that, given the exponential increase of relative errors with $t$,
the result of a fit to data at $t\geq t_{\min}$ will be almost identical
to the effective mass calculated at $t=t_{\min}$, we decided
to estimate the
potential by the latter. We also demanded
the effective mass at $t_{\min}-a$ to be compatible with the quoted
result, determined at
$t=t_{\min}$.

\begin{figure}
\centerline{\epsfxsize=8cm\epsfbox{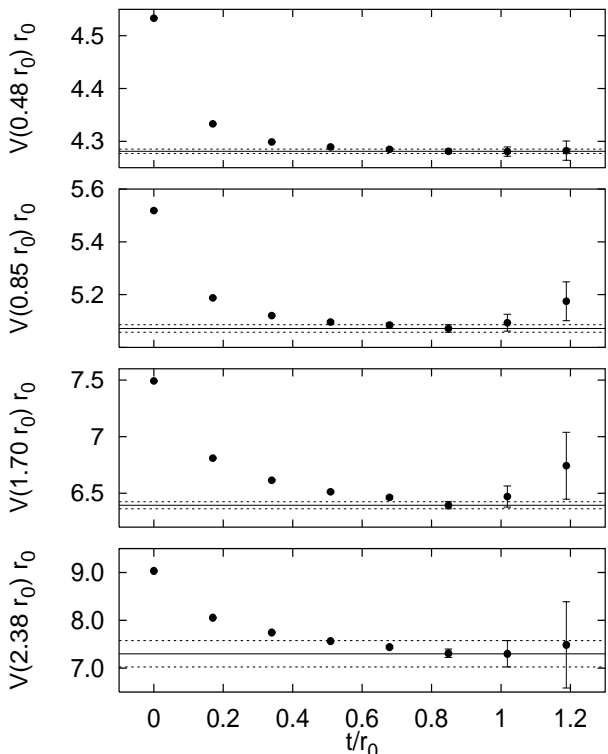}}
\caption{Effective masses
for the potential on the $24^3$ lattice at $\kappa=0.1575$
at four different source separations.}
\label{fig:effpot}
\end{figure}

In Fig.~\ref{fig:effpot}, we illustrate the convergence of effective
masses,
\begin{equation}
V({\mathbf r},t)=a^{-1}\ln\frac{W({\mathbf r},t)}{W({\mathbf r},t+a)}
\longrightarrow V({\mathbf r})\quad (t\rightarrow\infty),
\end{equation}
towards the asymptotic value, $V({\mathbf r})$, for the case of the
$\kappa = 0.1575$ potential on the $24^3$ volume at four randomly
selected distances. Note that, due to the positivity of the Wilson
action, the approach towards the plateau has to be monotonous.
At $\beta=5.6$, with the smearing algorithm used,
the range of $\kappa$ values investigated and the selection criterium
employed, we found $V[{\mathbf r},t_{\min}({\mathbf r})]$
to approximate the asymptotic value
within statistical errors for $4a\leq t_{\min}\leq 6a$, i.e.\ at
$t_{\min}\approx r_0$. In the case of the $\Pi_u$ potentials,
the ground state overlaps were found to be somewhat inferior and the
statistical errors larger. As a consequence of these two competing
effects
the necessary $t$-cuts came out to be almost identical to those
employed for the ground state potential.

For distances bigger than about $2\,r_0$, $t_{\min}$ had to be increased
by one lattice unit with respect to its value at smaller distances.
This differs from our experience from quenched
studies employing similar methods
at similar lattice spacings where systematic and statistical
effects happened to interfere in such a way that
the same $t$-cut could be applied at all
distances~\cite{Bali:1992ab,Bali:1993ru}.
Once the potential is extracted, the overlap of the creation operator
with the physical ground state can be quantified in terms of an
overlap coefficient, $0<  C_0({\mathbf r})\leq 1$:
\begin{equation}
C_0({\mathbf r},t)=W({\mathbf r},t)e^{V({\mathbf r},t)t}
\longrightarrow C_0({\mathbf r})\quad (t\rightarrow\infty).
\end{equation}

\begin{figure}
\centerline{\epsfxsize=8cm\epsfbox{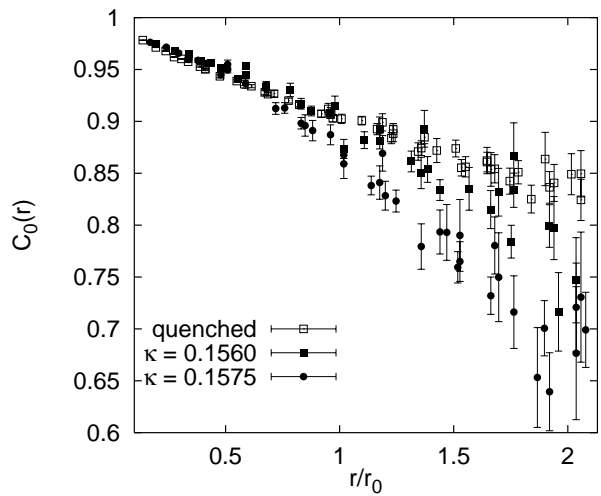}}
\caption{Ground state overlaps of the potential creation operators
as a function of $r$. The quenched results have been obtained at
$\beta=6.2$.}
\label{fig:over}
\end{figure}

In Fig.~\ref{fig:over}, we compare the ground state overlaps that have been
achieved
at $\kappa=0.156$ and $\kappa=0.1575$ ($L_{\sigma} =24$)
with quenched $\beta=6.2$ data that have been obtained by use
of the same smearing algorithm and analysis procedure.
We note that while at small $r$ the
ground state overlaps of the un-quenched simulations are superior to the
quenched reference data, at large $r$ the projection onto the physical
ground state becomes increasingly worse, in particular at the lighter
quark mass. 

The dependence of the quenched overlaps on
$r$ is linear to first approximation, however, this
is not so for the sea quark data (which explains why $t_{\min}$ had to be
increased as a function of $r$ in the latter case). We conclude that
at large $r$ the flux tube distribution appears to
change when including sea quarks. This effect, which becomes more pronounced
when the quark mass is decreased, can be interpreted as
a first indication of string breaking: the physical ground
state will be
a mixture between a would-be static-static state and a
pair of two static-light mesons. While our creation operator is
tuned to optimally project onto the former it has almost zero overlap with a
wave function of the latter type.
As a consequence, we lose local mass plateaus altogether
in the regime, $r>2.5\, r_0$. A similar reduction of
ground state overlaps at large distances has been reported by the 
CP-PACS Collaboration~\cite{Aoki:1999ff}.

\section{Results}
\label{sec:results}
\subsection{The static potentials}
\label{sec:pot}
The potentials obtained from the six
dynamical fermion and two quenched simulations have been rescaled 
in units of $r_0$. Subsequently, the constant $r_0V(r_0)$
has been subtracted in all cases to cancel the (different)
static self energy contributions and to achieve the common normalisation,
$V(r_0)=0$.
Up to violations of rotational symmetry at small distances,
the un-quenched potentials are found to agree with each other within
statistical errors.
In particular, we do not see finite size effects
among the two $\kappa=0.1575$ data sets.

\begin{figure}
\centerline{\epsfxsize=8cm\epsfbox{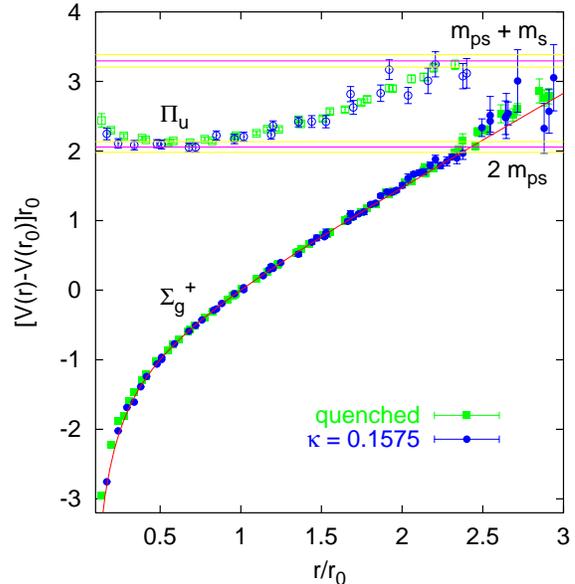}}
\caption{Comparison between quenched ($\beta=6.2$) and un-quenched
ground state and $\Pi_u$ potentials.}
%\vskip -1cm
\label{fig:pot}
\end{figure}

In Fig.~\ref{fig:pot}, we compare
the potential obtained on the largest lattice volume at our disposal,
$L_{\sigma}a=24\,a\approx 4.07\,r_0\approx 2.03$~fm at $\kappa=0.1575$,
with the quenched potential at $\beta=6.2$. Besides the
ground state potentials, which correspond to the $\Sigma_g^+$ representation
of the relevant continuum symmetry group $D_{\infty h}$,
the $\Pi_u$ hybrid potentials~\cite{Campbell:1984fe,Bali:2000gf}
are depicted as well as the approximate
masses
of pairs of
static-light scalar and pseudoscalar bound states into which the 
static-static string states
are expected to decay at large $r$.
Note that for separations, $r>1.9\,r_0$, we have used some extra
off-axis directions, in addition to those mentioned in Sect.~\ref{sec:lat}.
Around $r\approx 2.3\,r_0$ we
expect both, $\Sigma_g^+$ and $\Pi_u$ potentials to exhibit
colour screening.
However, the Wilson loop
data are not yet precise enough to resolve this effect.
In a forthcoming paper~\cite{inprep} we therefore
intend to present a more systematic analysis geared to improve
the statistical quality of
potential data in the interesting regime, $r>2\,r_0$, along with a
study of the corresponding transition matrix elements that might shed more
light onto this question~\cite{Pennanen:2000yk,DeTar:1999ie}.

\begin{figure}
\centerline{\epsfxsize=8cm\epsfbox{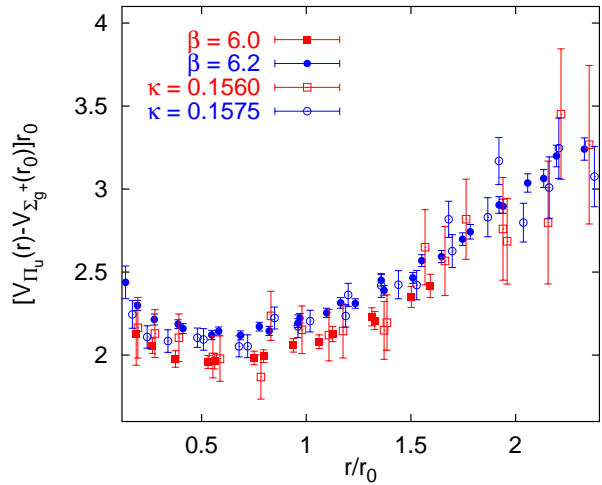}}
\caption{First hybrid excitation: quenched vs.\ un-quenched.}
\label{fig:hybrid}
\end{figure}

In Fig.~\ref{fig:hybrid}, we compare our two most precise
un-quenched data sets ($\kappa=0.156$ and $\kappa=0.1575$)
on the hybrid potential with quenched results.
We find no
statistically significant differences, other than
finite $a$ effects between the quenched potentials obtained
at two different $\beta$ values (full symbols).

The ground state potentials are fitted to the parametrisation,
\begin{equation}
\label{eq:fit}
V({\mathbf r})=V_0+\sigma\,r-\frac{e}{r}+g
\left(\frac{1}{r}-\left[\frac{1}{\mathbf r}\right]\right),
\end{equation}
where the term $[1/{\mathbf r}]$, that denotes the tree level lattice
propagator in position space~\cite{Bali:2000gf}, is included
to quantify the short distance
lattice artefacts. Given the fact that the number of elements
in the covariance matrix is much bigger than the sizes of our
data samples, we refrain from attempting correlated fits.
Therefore, $\chi^2$ can in principle become much smaller than the
degrees of freedom, $N_{DF}$. All errors are bootstrapped.

We wish to investigate the effect of sea quarks onto the short range interquark
potential.
Apart from the fit parameters, we also calculate the Sommer scale
of Eq.~(\ref{eq:som}),
\begin{equation}
r_0=\sqrt{\frac{1.65-e}{\sigma}}.
\end{equation}
The fit function, Eq.~(\ref{eq:fit}), is only thought to effectively
parameterise the potential within a given distance regime, rather than
being theoretically sound. For instance neither string breaking 
nor the running of the QCD coupling has
been incorporated.
Therefore, the values of the effective fit parameters
will in general
depend on the fit range, $[r_{\min},r_{\max}]$. For instance,
as a consequence of asymptotic freedom,
$e$ will weaken as data points from smaller and smaller
distances are included. To exclude such a systematic bias from our
comparison between quenched and un-quenched fit parameters, we deviate
somewhat from our first analysis~\cite{Glassner:1996xi}
(as we have already done in Refs.~\cite{Bali:1998bj,Gusken:1998sa})
and use the same $r_{\min}>0.4\,r_0$ in physical units for all our fits
and $r_{\max}<aL_{\sigma}/2$. The results are
displayed in Tab.~\ref{tab:pot}.

Three-parameter
fits have been performed
in addition, by constraining $g=0$ in Eq.~(\ref{eq:fit}).
To obtain acceptable fit qualities for the same physical fit range
on all data sets we had to increase, $r_{\min}\approx 0.6\,r_0$.
The latter fit results are shown in Tab.~\ref{tab:pot2}.

\begin{table}
\caption{Four-parameter fits to the static potential for $r_{\min}>0.4\,r_0$.}
\label{tab:pot}
\begin{tabular}{cccccccc}
$\kappa$&$L_{\sigma}$&$\frac{r_{\min}}{a}$&$\frac{\chi^2}{N_{DF}}$&$g$&$V_0a$
&$e$&$\sqrt{\sigma}a$\\
\hline
0.1560&16&$\sqrt{5}$&$0.94$&0.27(5)&0.731(06)&0.338(09)&0.224(2)\\
0.1565&16&$\sqrt{5}$&$0.54$&0.42(7)&0.738(10)&0.349(16)&0.216(3)\\
0.1570&16&$\sqrt{5}$&$0.78$&0.34(7)&0.725(12)&0.324(19)&0.210(4)\\
0.1575&16&$\sqrt{6}$&$0.47$&0.30(9)&0.739(11)&0.337(20)&0.192(4)\\
0.1575&24&$\sqrt{6}$&$0.76$&0.30(6)&0.746(05)&0.352(09)&0.193(2)\\
0.1580&24&$2\sqrt{2}$&$0.64$&0.28(7)&0.751(07)&0.358(12)&0.182(3)\\
$\kappa_{ph}$
& --- & --- & --- & --- &0.760(12)&$0.368^{+20}_{-26}$&
$0.171^{+6}_{-3}$\\\hline
$\beta=6.0$&16&$\sqrt{5}$&$0.14$&0.34(2)&0.658(04)&0.291(05)&0.219(2)\\
$\beta=6.2$&32&$3$&$0.49$&0.25(1)&0.636(03)&0.292(05)&0.160(1)
\end{tabular}
\end{table}

\begin{figure}
\centerline{\epsfxsize=8cm\epsfbox{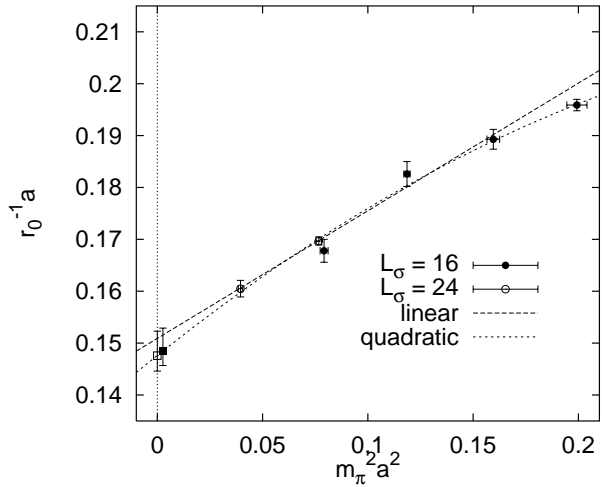}}
\caption{Chiral extrapolation of the scale $r_0^{-1}$.
The open square corresponds to the chiral limit, the full square
to the physical ratio, $r_0m_{\pi}\approx 0.350$.}
\label{fig:v0}
\end{figure}

In Tab.~\ref{tab:r0}, we display the $r_0$ values obtained from
our four-parameter interpolations as well as the effective string tension in
units of $r_0$. The same data is depicted in Fig.~\ref{fig:v0} as
a function of the squared $\pi$ mass, $(m_{\pi}a)^2$.
Chiral extrapolations are performed according to,
\begin{equation}
a\,r_0^{-1}(m_{\pi})=a\,r_0^{-1}(0)+c_1a^2\,m_{\pi}^2+c_2a^4\,m_{\pi}^4.
\end{equation}
The fit to all data points, with the exception of the
$L_{\sigma}=16$ result at $\kappa=0.1575$, yields the parameter values,
\begin{eqnarray}
a\,r_0^{-1}(0)&=&0.1476(30),\\
c_1&=&0.324(57),\\
c_2&=&-0.41(22),\\
\chi^2/N_{DF}&=&1.37/2.
\end{eqnarray}
When excluding the rightmost point ($\kappa=0.156$) a linear extrapolation
becomes possible:
\begin{eqnarray}
a\,r_0^{-1}(0)&=&0.1509(14),\\
c_1&=&0.246(16),\\
\chi^2/N_{DF}&=&1.32/2.
\end{eqnarray}
The fitted curves as well as the quadratically extrapolated
$r_0^{-1}(0)$ value (open square), with error bars enlarged to incorporate
the one $\sigma$ region around
the linear extrapolation, are included in the Figure.

\begin{table}
\caption{Three-parameter fits to the static potential for $r_{\min}\approx
0.6\,r_0$.}
\label{tab:pot2}
\begin{tabular}{ccccccc}
$\kappa$&$L_{\sigma}$&$\frac{r_{\min}}{a}$&$\frac{\chi^2}{N_{DF}}$&$V_0a$&$e$&
$\sqrt{\sigma}a$\\
\hline
0.1560&16&$3$&$0.84$&0.712(12)&0.290(23)&0.228(3)\\
0.1565&16&3&$0.48$&0.715(21)&0.291(42)&0.221(6)\\
0.1570&16&$3$&$1.43$&0.725(21)&0.326(42)&0.210(6)\\
0.1575&16&$3$&$0.57$&0.727(19)&0.308(38)&0.195(5)\\
0.1575&24&$3$&$0.75$&0.736(07)&0.322(15)&0.195(2)\\
0.1580&24&$3$&$0.77$&0.747(08)&0.348(17)&0.183(3)\\\hline
$\beta=6.0$&16&$3$&$0.87$&0.639(06)&0.245(19)&0.223(3)\\
$\beta=6.2$&32&$3\sqrt{2}$&$0.67$&0.637(05)&0.296(13)&0.160(1)
\end{tabular}
\end{table}

\begin{table}
\caption{$r_0$ and the ratio $\sqrt{\sigma}r_0$.}
\label{tab:r0}
\begin{tabular}{cccc}
$\kappa$&$L_{\sigma}$&$r_0/a$&$\sqrt{\sigma}r_0$\\\hline
0.1560&16&5.104(29)&1.145(4)\\
0.1565&16&5.283(52)&1.141(7)\\
0.1570&16&5.475(72)&1.152(8)\\
0.1575&16&5.959(77)&1.146(9)\\
0.1575&24&5.892(27)&1.139(4)\\
0.1580&24&6.230(60)&1.137(5)\\
$\kappa_{ph}$& --- &$6.73^{+13}_{-19}$&$1.133_{-8}^{+11}$\\\hline
$\beta=6.0$&16&5.328(31)&1.166(2)\\
$\beta=6.2$&32&7.290(34)&1.165(2)
\end{tabular}
\end{table}

An extrapolation to the physical limit,
$m_{\pi_{ph}}r_0\approx 0.350$, yields,
\begin{equation}
a\,r^{-1}_0(m_{\pi_{ph}})=0.1485^{+44}_{-28}.
\end{equation}
The corresponding $r_0/a$ value is included in Tab.~\ref{tab:r0}.
Note that $\kappa_{ph}=0.158456(19)$ only marginally differs
from $\kappa_c=0.158493(18)$.

From bottomonium phenomenology~\cite{Bali:2000gf,Sommer:1994ce,Bali:1997am}
one obtains, $r_0^{-1}=(394\pm 20)$~MeV. The lattice spacing
determined from  $r_0$ at physical sea quark mass, therefore, is,
\begin{equation}
a^{-1}=\left(2.68^{+8}_{-11}\pm 0.14\right)\mbox{~GeV}.
\end{equation}
The last error reflects the scale
uncertainty within the phenomenological $r_0$ determination.
The above value compares
well with $a^{-1}=2.7(2)$~GeV,
as obtained from the $\rho$ mass~\cite{inprep2,Eicker:1999sy}
after an extrapolation to
the physical $m_{\pi}/m_{\rho}$ ratio.
By interpolating between quenched reference data~\cite{Bali:2000gf},
we find that the above $r_0$ value corresponds to the quenched
$\beta=6.14(2)$: inclusion of two light Wilson quarks at $\beta=5.6$ thus results
in a $\beta$ shift,
$\Delta\beta=0.54(2)$, i.e.\ in an increase of the lattice
coupling, $g^2=6/\beta$, by 9--10 \%.

\begin{figure}
\centerline{\epsfxsize=8cm\epsfbox{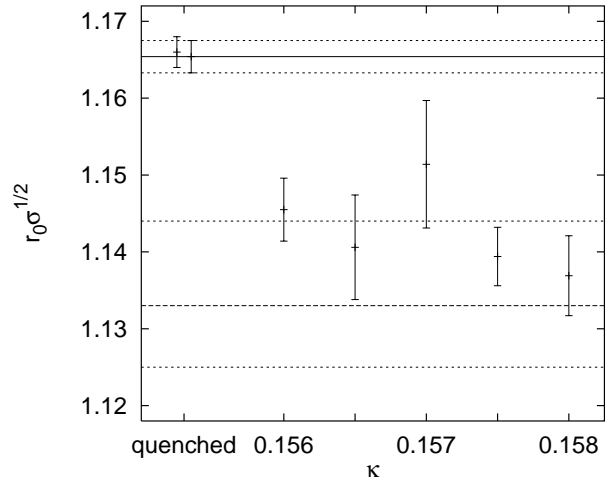}}
\caption{Sea quark mass dependence of the ratio between the
square root of the string tension and $r_0^{-1}$. The horizontal lines with
error bands correspond to the quenched (solid) and
chirally extrapolated un-quenched (dashed) results.}
\label{fig:string}
\end{figure}

\begin{figure}
\centerline{\epsfxsize=8cm\epsfbox{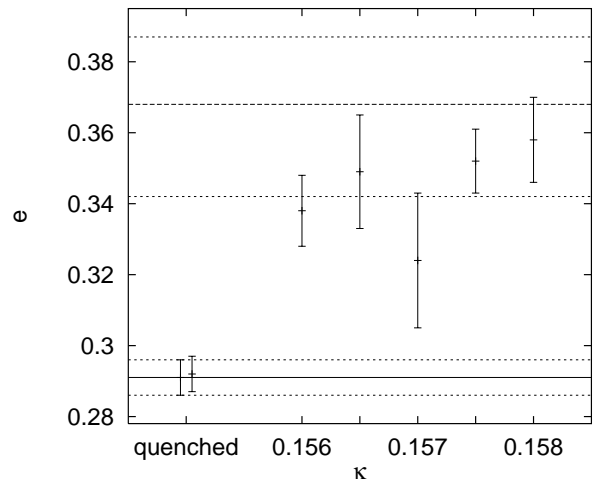}}
\caption{Sea quark mass dependence of the effective Coulomb strength
of the static potential.}
\label{fig:coul}
\end{figure}

The values of
$V_0a$, $\sqrt{\sigma}a$ and $e$, extrapolated to the physical quark mass,
are displayed in Tab.~\ref{tab:pot} while the combination
$\sqrt{\sigma}r_0$ is included in Tab.~\ref{tab:r0}.
In Figs.~\ref{fig:string} and \ref{fig:coul},
we display $\sqrt{\sigma} r_0$ and $e$ against $\kappa$, respectively.
In these two cases not only quadratic and linear fits in $(m_{\pi}a)^2$ but
also simple averages
have been taken into account when assigning the error bars, that
reflect both statistical errors as well as the uncertainty in the
parametrisation, to the
(quadratically) extrapolated values.

From Fig.~\ref{fig:string} as well as from Tab.~\ref{tab:r0} we can
read off that the ratio $\sqrt{\sigma}r_0$ decreases down to
$1.133^{+11}_{-8}$, compared to the quenched result, 1.165(3).
With $r_0\approx 0.5$~fm
we obtain a quenched value, $\sqrt{\sigma}\approx 460$~MeV, while
with two sea quarks we find, $\sqrt{\sigma}\approx 445$~MeV.
Note that
the latter result comes closer to estimates, $\sqrt{\sigma}=(429\pm 2)$~MeV,
from the $\rho,a_2,\ldots$ Regge trajectory~\cite{Bali:2000gf}.

We find self energy, $V_0a$, and effective
Coulomb coefficient, $e$, to significantly increase with respect to
the quenched simulations. In tree level perturbation theory
both parameters, $V_0a$ and $e$, are expected to be proportional to
$g^2$. Data on $e$
are shown in Fig.~\ref{fig:coul}.
The results, extrapolated to the physical point,
are (Tab.~\ref{tab:pot}),
$e=0.368^{+20}_{-26}$ and $V_0a=0.760(20)$,
i.e.\ $e$ is increased by
16 -- 33 \% and $V_0$ by 16 -- 21 \%
with respect to the quenched results: the effect of sea quarks
on couplings defined through the potential at short range is bigger than
the relative shift in the lattice coupling, $\Delta\beta/\beta\approx 0.1$.
Bottomonia spectroscopy suggests a value~\cite{Bali:1999pi}, $e\approx 0.4$,
which, given our $n_f=2$ result, is indeed likely
to be consistent with
QCD with three light sea quark flavours.
The three-parameter fits (Tab.~\ref{tab:pot2}) are, due to the
different fit range
$r\geq 0.6\,r_0$, not yet precise enough to confirm an increase of $e$
beyond doubt. However, at least
the effect on $V_0a$ is statistically significant and in agreement with
the result from the four-parameter fits.

\begin{figure}
\centerline{\epsfxsize=8cm\epsfbox{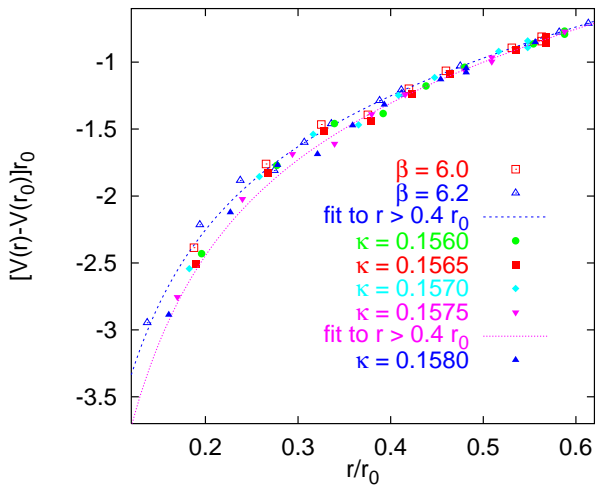}}
\caption{The short distance potential: quenched (open symbols)
vs.\ un-quenched (full symbols).}
\label{fig:potshort}
\end{figure}

In Fig.~\ref{fig:potshort}, we compare quenched (open symbols)
and un-quenched (full symbols) lattice data
at short distances to visualise that our interpretation
of an increase in the
effective Coulomb strength is independent of the
fitted parametrisation.
The curves correspond to
the values of the parameters $\sigma r_0^2$, $e$ and $V_0r_0$
at $\beta=6.2$
(quenched) and $\kappa=0.1575$ (un-quenched)
that have been obtained in the four-parameter fits. Due to lattice
artefacts the data sets scatter around the interpolating curves.
As expected,
we indeed observe that the un-quenched data points (full symbols)
lie systematically below
their quenched counterparts (open symbols). 

\subsection{Torelons}
\label{sec:tor}
In view of the difficulties in detecting string breaking directly in the
static potential, it might be illuminating to investigate the effect
of including sea quarks on torelon
masses~\cite{Michael:1989vh}. Such numerical
simulations have been pioneered
by Kripfganz and Michael~\cite{Kripfganz:1989jv}
on small lattice volumes some ten years ago.
In pure gauge theories
a torelon that winds around a dimension of extent $L_{\sigma} a$
will have a mass,
\begin{equation}
\label{eq:tor1}
m_T=aL_{\sigma}\left[\sigma+\Delta\sigma(aL_{\sigma})\right],
\end{equation}
in the limit of large volumes, $aL_{\sigma}\rightarrow\infty$. 
$\sigma$ should be the very same string tension
that governs the static potential at large distances. 
The sub-leading finite size correction~\cite{Ambjorn:1984yu},
\begin{equation}
\label{eq:tor2}
\Delta\sigma=-\frac{\pi}{3(aL_{\sigma})^2},
\end{equation}
is expected from the bosonic string picture.
The effect of
this contribution,
which has been accurately verified in $SU(2)$ gauge
theory~\cite{Michael:1994ej}, ranges from a
decrease by 11~\% on our smallest physical volume
($16^3$ at $\kappa=0.1575$)
to 4.9~\% on our biggest volume ($24^3$ at $\kappa=0.1575$).

In a pure gauge theory the centre symmetry of the action implies
that all torelons that wind once around  a lattice boundary are
mass degenerate (see e.g.\ 
Appendix D.2 of Ref.~\cite{Bali:2000gf} for details).
When explicitly violating the centre symmetry by
including sea quarks the degeneracy will be broken,
with torelons to be classified in accord to representations
of the cubic group,
$O_h$ times charge conjugation. While in pure gauge theories
glueballs can only split up into pairs of torelons, with sea quarks
a single torelon can mix with or decay into a glueball that is in the
same representation.
For this to happen the torelon has to break up, unwind and rearrange
itself into
a flux loop with trivial winding number.
In reality, one will meet
different states (and creation operators)
with equal quantum numbers. The operator
that corresponds to the torelon in the quenched case will
have maximal overlap with the state that is most torelon like but will
in general, in the limit
of large Euclidean times, decay into
the lightest possible
state. In the case that another state turns out to be lighter than
the expected torelon mass we
would call this effect ``string breaking'' of the torelon.

Heuristically one would expect a torelon-type operator to project better
onto a state with a large
mesonic component than a state with a large
glueball component since
the former appears as an intermediate state in the deformation
of a torelon into a glueball anyway.
On the other hand progress is being made in the excitation
of flavour singlet states by use
of mesonic operators (quark loops)~\cite{AliKhan:1999zi,Michael:1999rs,Sesam:2000}
and there are indications~\cite{Michael:1999rs} that purely gluonic operators
of the type we use as well as mesonic operators might both have
acceptable overlaps with the physical ground state.

\begin{table}
\caption{Torelon masses.}
\label{tab:torelon}
\begin{tabular}{ccccccc}
$\kappa$&$L_{\sigma}$&
$m_{A_1^{++}}r_0$&$m_{E^{++}}r_0$&$m_{T_1^{+-}}r_0$&expected&
$2m_{\pi}r_0$\\\hline
0.1560&16&3.98(30)&4.03(20)&3.84(32)&3.78(07)&4.56(4)\\
0.1565&16&3.34(36)&3.43(25)&3.59(42)&3.59(11)&4.22(6)\\
0.1570&16&3.53(25)&3.54(19)&3.06(33)&3.51(14)&3.77(6)\\
0.1575&16&2.85(17)&2.82(15)&2.82(18)&3.11(12)&3.32(5)\\
0.1575&24&4.41(68)&6.16(66)&4.06(63)&5.03(08)&3.26(3)\\
0.1580&24&6.11(93)&5.36(44)&2.24(98)&4.71(13)&2.47(5)    
\end{tabular}
\end{table}

\begin{figure}
\centerline{\epsfxsize=8cm\epsfbox{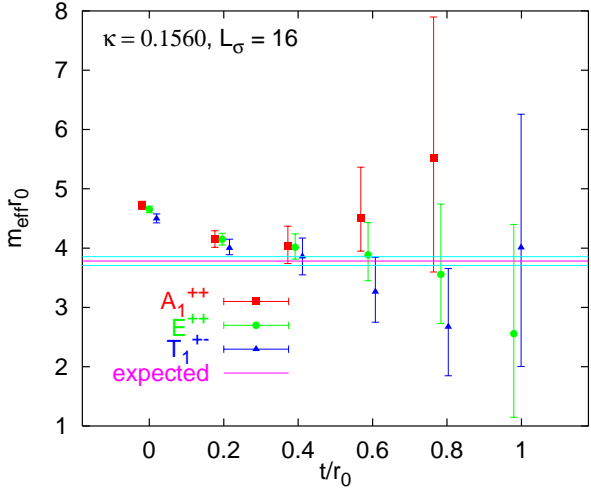}}
\caption{Effective torelon masses at $\kappa=0.156$,
in comparison with the expectation,
Eqs.~(\protect\ref{eq:tor1}) -- (\protect\ref{eq:tor2}), with
the effective
string tension $\sigma$ obtained from a fit to the static potential.}
\label{fig:tor1}
\end{figure}

\begin{figure}
\centerline{\epsfxsize=8cm\epsfbox{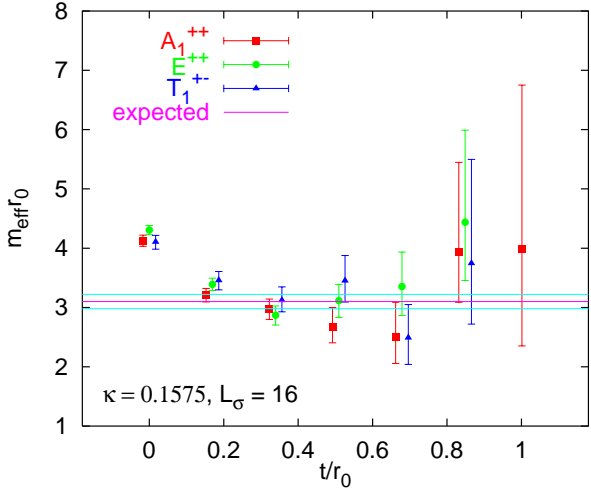}}
\caption{Same as Fig.~\protect\ref{fig:tor1} at $\kappa=0.1575$.}
\label{fig:tor3}
\end{figure}

In addition to
the fitted torelon masses, we display
the non-string breaking expectations of Eqs.~(\ref{eq:tor1})
-- (\ref{eq:tor2}) in Tab.~\ref{tab:torelon},
based on the string tensions from our
four-parameter fits to the static potential of Tab.~\ref{tab:pot}.
Unfortunately, the signals on
the large lattices become very noisy,
due to the larger torelon mass.
Apart from isoscalar meson/glueball states,
the $A_1^{++}$ torelon can
in principle decay into a pair of $\pi$'s. We therefore
include for guidance twice the
$\pi$ mass in the last column of the Table.
As we shall see below (Tab.~\ref{tab:glue}),
on all $16^3$ lattices, the scalar
glueball comes out to have a mass very close to the torelon mass
expectation of Eqs.~(\ref{eq:tor1}) and (\ref{eq:tor2}).
On our $24^3$ lattices, however, the expectation
supercedes the $0^{++}$
and, eventually, the $1^{+-}$ glueball masses:
breaking of the $A_1^{++}$ and, eventually, $T_1^{+-}$ torelons
into a flavour singlet meson/glueball state becomes energetically possible.
At the same time two $\pi$'s drop below the torelon expectation
(and become lighter than the flavour singlet meson/glueball state).

\begin{figure}
\centerline{\epsfxsize=8cm\epsfbox{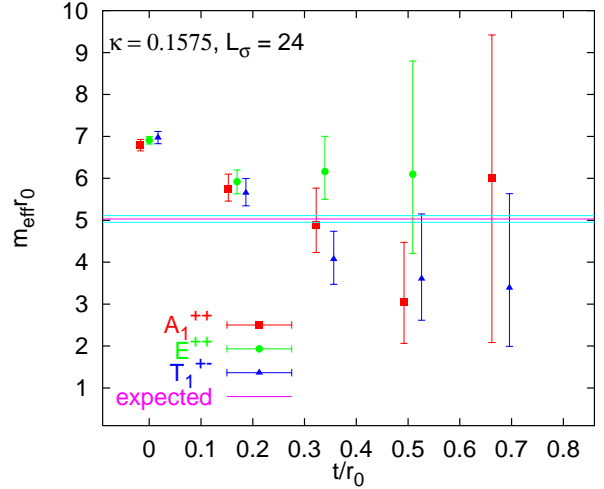}}
\caption{Same as Fig.~\protect\ref{fig:tor3} on the larger volume.}
\label{fig:tor4}
\end{figure}

\begin{figure}
\centerline{\epsfxsize=8cm\epsfbox{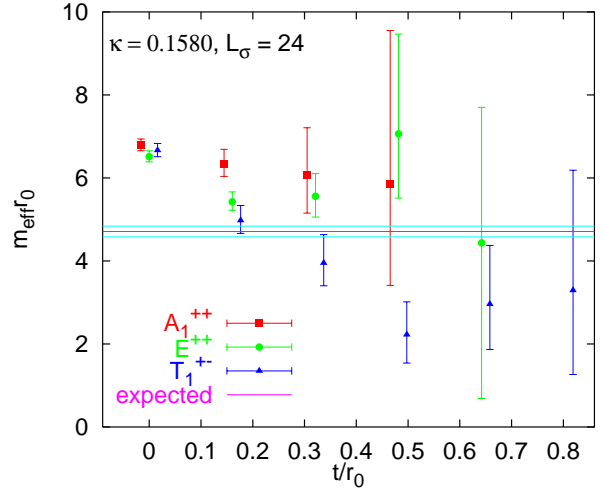}}
\caption{Same as Fig.~\protect\ref{fig:tor4} at $\kappa=0.158$.}
\label{fig:tor5}
\end{figure}

On all the $16^3$ lattices, within statistical errors,
the three torelons are mass degenerate
and in agreement with the non-string breaking
expectation. Two effective mass plots
are displayed as examples in Figs.~\ref{fig:tor1} and \ref{fig:tor3}.
The same holds true, within larger errors, in the case of the
$24^3$ simulation at $\kappa=0.1575$ (Fig.~\ref{fig:tor4}).
However, at $\kappa=0.158$ with a mass of $(2.24\pm 0.98)\,r_0$
at least the
$T_1^{+-}$ torelon appears to fall short of the expectation
by two and a half
standard deviations (Fig.~\ref{fig:tor5})! 
It would be important to increase statistics in order to
corroborate this result as a signal for string breaking.

\subsection{Glueballs}
\label{sec:glue}
We display the extracted masses of the three glueballs
that we investigated
in Tab.~\ref{tab:glue}. The quenched results at
a similar lattice spacing ($\beta=6.0$) have been calculated by Michael
and Teper\footnote{In the case of the $2^{++}$
glueball we display their $E^{++}$ result only since the $T_2^{++}$
representation has not been realised in our
study.}~\cite{Michael:1989jr}.
The quenched continuum
$0^{++}$ result~\cite{Bali:2000gf} (last row of the Table) stems from a
quadratic extrapolation in
the lattice spacing of data obtained in
Refs.~\cite{deForcrand:1985rs,Michael:1988wf,Michael:1989jr,Bali:1993fb,Vaccarino:1999ku} while the tensor and axialvector results are from
Morningstar and Peardon~\cite{Morningstar:1999rf}.

\begin{figure}
\centerline{\epsfxsize=8cm\epsfbox{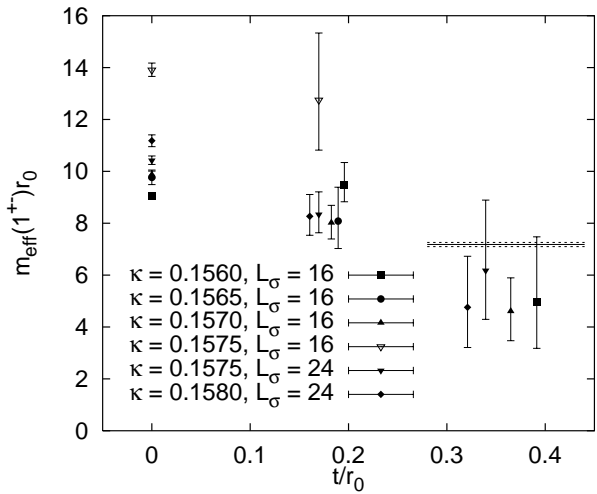}}
\caption{Effective $1^{+-}$ glueball masses. The horizontal
line corresponds to the quenched continuum result.}
\label{fig:axvec}
\end{figure}

\begin{figure}
\centerline{\epsfxsize=8cm\epsfbox{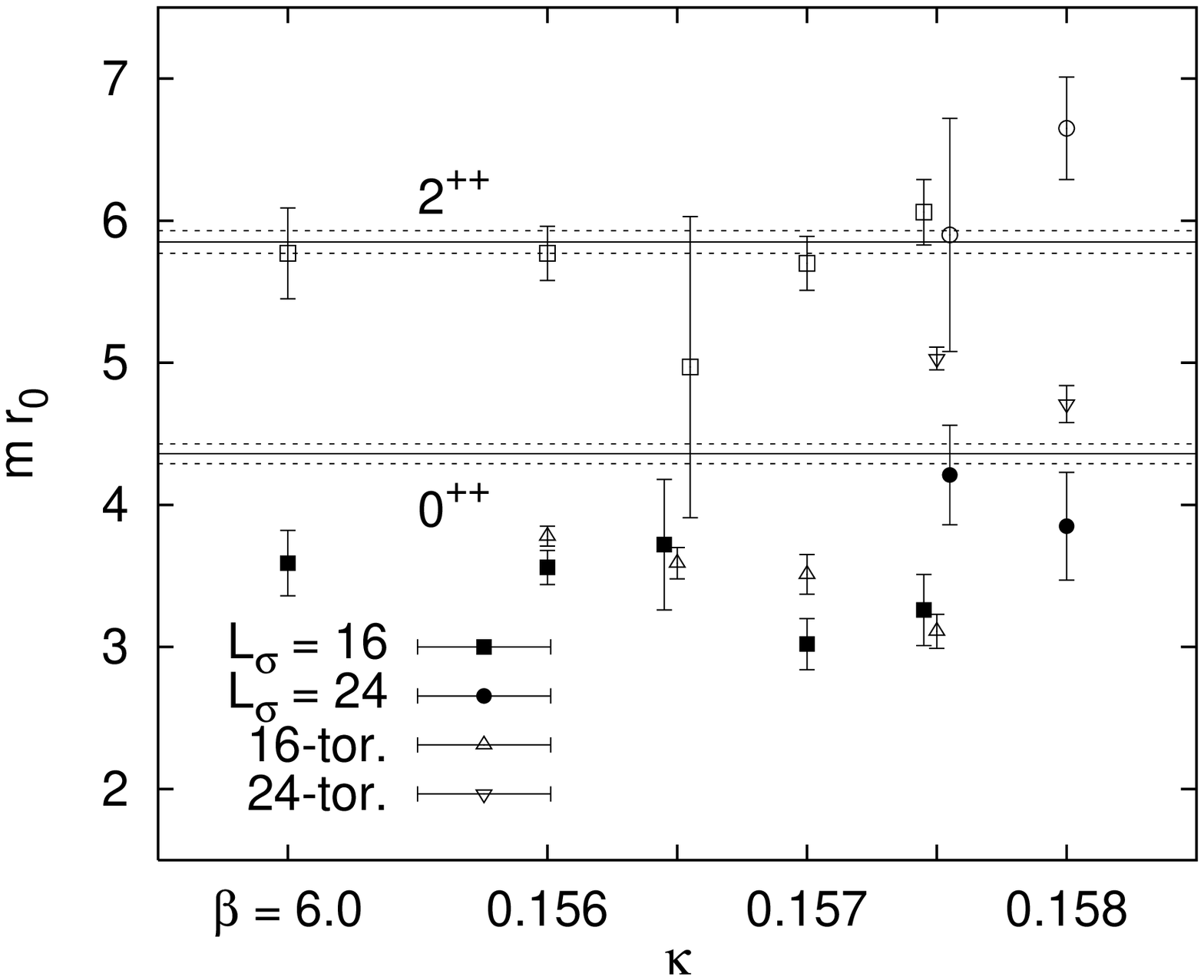}}
\caption{Results on scalar and tensor glueball masses.
The horizontal lines with error bands indicate quenched continuum
results while the leftmost data points
correspond to quenched Wilson action results obtained at 
finite lattice spacing ($\beta=6.0$). The expected torelon masses
on $16^3$ and $24^3$ lattices are shown too (open triangles).}
\label{fig:glueres}
\end{figure}

Effective mass plots of the scalar and tensor channels are shown in
Figs.~\ref{fig:glue1} -- \ref{fig:glue6}.
In the case of $J^{PC}=1^{+-}$ no plateau
has been established and the values displayed in the Table only represent
upper limits.
The corresponding effective masses are
displayed in Fig.~\ref{fig:axvec}, in comparison to the continuum
limit extrapolated quenched result (horizontal line with error band).
Quenched results obtained at similar lattice spacings~\cite{Michael:1989jr}
suggest
the glueball to lie
somewhere inbetween $6.5\,r_0^{-1}$ and $8.5\,r_0^{-1}$ and, within
statistical errors, do not deviate from the continuum value.
Keeping in mind that all un-quenched simulations
are statistically independent of each other, there is a clear indication
for a
$1^{+-}$ state being lighter than $6\,r_0^{-1}$.
It remains to be demonstrated in future simulations at
different lattice couplings, $\beta$, whether this conclusion persists
in the continuum limit.

In Fig.~\ref{fig:glueres}, our results on the scalar and tensor glueballs
are summarised. The leftmost data points represent the $\beta=6.0$
quenched results while the horizontal error bands correspond to the
respective quenched continuum limits. The quenched scalar
glueball at the finite lattice spacing comes out to be lighter by
15--20~\% than the continuum limit extrapolated
value~\cite{Bali:1993fb,Bali:2000gf} while the tensor glueball is in agreement
with the continuum result.
In addition to the glueball masses,
the torelon mass expectations of
Eqs.~(\ref{eq:tor1}) -- (\ref{eq:tor2}) are included into the Figure.
For $\kappa\leq 0.157$, $2m_{\pi}$ is larger than the torelon mass.
On the $16^3$ lattice at $\kappa=0.1575$ the two masses come
out to be degenerate within errors
and on both $24^3$ lattices two
$\pi$'s become lighter than the torelon (and the scalar glueball).

\begin{table}
\caption{Glueball masses. A star
indicates that no effective mass plateau could be identified
and, therefore, only an upper limit is stated.}
\label{tab:glue}
\begin{tabular}{cccccc}
$\kappa$&$L_{\sigma}$&$m_{0^{++}}r_0$&$m_{2^{++}}r_0$&$m_{1^{+-}}r_0$&
$m_{2^{++}}/m_{0^{++}}$\\
\hline
0.1560&16&3.56(12)&5.77(19)&5.0(2.1)*&1.62(07)\\
0.1565&16&3.72(46)&4.97(106)&8.1(1.2)*&1.34(34)\\
0.1570&16&3.02(18)&5.70(19)&4.6(1.2)*&1.89(13)\\
0.1575&16&3.26(25)&6.06(23)&12.7(2.3)*&1.86(13)\\
0.1575&24&4.21(35)&5.90(82)&6.2(2.3)*&1.41(23)\\
0.1580&24&3.85(38)&6.65(36)&4.8(1.8)*&1.72(17)\\\hline
$\beta=6.0$&16&3.57(21)&5.70(32)&7.5(1.1)&1.60(14)\\
$\beta=\infty$& --- &4.36(7)&5.85(8)&7.18(8)&1.34(3)\\
\end{tabular}
\end{table}

We do not detect any statistically
significant deviation between quenched and un-quenched results
in the $2^{++}$ channel. Unfortunately, the statistical error
on the $24^3$ lattice at $\kappa=0.1575$ is too large to resolve possible
finite size effects.
We find all $16^3$ $0^{++}$ glueball data to roughly agree with
the quenched $\beta=6.0$ result. At the same time, within
statistical errors,
the scalar glueball
masses are degenerate with the $A_1^{++}$ (expected and measured)
torelon masses. This degeneracy
has also been observed in Ref.~\cite{Kripfganz:1989jv},
on small lattice volumes: in the quenched approximation,
finite size effects are small as long as two torelons are heavier than
the glueball, however, with sea quarks, there is no protection preventing
the glueball
to decay into a single torelon.
This could explain why the
$0^{++}$ glueball on the 2~fm lattice at $\kappa=0.1575$
comes out to be significantly heavier
than the one on the 1.4~fm lattice. 
Note that the size of the corresponding wave function becomes
reduced when increasing the lattice volume (Sect.~\ref{sec:mass}).

On both $24^3$ lattices
the mass of the scalar glueball comes out to be bigger than
$2m_{\pi}$. There is no physical reason
why the creation operator employed should have zero overlap
with either a pair of $\pi$'s or
a hypothetical $\pi\pi$ bound state~\cite{Alford:2000mm}.
However, it seems that
this overlap is small, similar to
the situation of the static potential, in which
the smeared Wilson loop only receives a tiny
contribution from pairs of static-light states.
In conclusion, the mass of the state onto which the scalar
glueball
operator dominantly projects
is compatible with quenched glueball results obtained at $\beta\geq 6$.
We find finite size effects that can be interpreted as
mixing with torelon states.

\section{Summary and Discussion}
We have determined various ``pure gauge'' quantities in a simulation
of QCD. Including two sea quark flavours results
in a shift of about 10~\% in $\beta$ and in an increase of the effective
Coulomb strength, governing the interquark potential
at $r>0.2$~fm, by 16 -- 33~\%, in the limit of light quark masses.
While we have not been able to establish string breaking
either in the ground state or in the $\Pi_u$ hybrid
potential, a comparison of the ground state overlaps
with quenched reference data
suggests an effect at large distances.

Around $r_c\approx 2.3\,r_0\approx 1.15$~fm
string breaking
will become energetically possible for both potentials,
ground state and hybrid, at $\kappa=0.1575$.
On the assumption that reducing the bare quark mass,
$m=(\kappa^{-1}-\kappa_c^{-1})/(2a)$, by an amount, $\Delta m$,
induces an equal change in the static-light meson mass,
we would guesstimate the string breaking length scale to drop by
$\Delta r\approx 2\Delta m/\sigma\approx 0.18\,r_0$ when extrapolating
to the physical sea quark mass.
This crude estimate tells us that with two light quark flavours the
QCD potential should become flat around $r_c\approx 1.05$~fm, such that in
``real'' $2+1$ flavour QCD string breaking at distances of 1~fm or
smaller is likely.

For $\kappa\leq 0.1575$ torelon masses are in agreement with the
(non-string breaking) expectation in terms of the
effective string tension, obtained from
a fit to the static potential, while at $\kappa=0.158$ there are indications
for the torelon in one particular representation to become lighter than
one would na\"\i{}vely have expected. The basic problem is that 
we are only faced with a small
window of lattice volumes for observation of ``string breaking''.
On a large lattice torelons become so heavy that
the correlation functions
disappear into noise at small temporal separation while on a
small lattice the
expected mass does not yet sufficiently differ
from the masses of decay candidates
to resolve an effect.
It would be helpful in this respect to have for instance
$\kappa=0.1575$ data at
$L_{\sigma}=20$.
The tensor glueball comes out to be heavier
than the respective torelon state, i.e.\ in this case
the two creation operators have little overlap with each other.

On our largest lattice ($\kappa=0.1575$) that corresponds to
$m_{\pi}/m_{\rho}=0.704(5)$ with extent,
$L_{\sigma}a\approx 4.07\,r_0$, we find,
$m_{0^{++}}=4.21(35)\,r_0^{-1}$, which is in agreement with the
quenched continuum limit result, $m_{0^{++}}=4.36(7)\,r_0^{-1}$.
The $2^{++}$ glueball masses also seem to agree with the quenched results
while there are indications for a light $T_1^{+-}$ glueball.
On small lattices we find the scalar glueball and
torelon to be degenerate and the glueball
mass is underestimated.
To exclude such finite size effects, the spatial lattice extent
should be taken to be bigger than about $3.5\,r_0\approx 1.75$~fm.

Previous results with two flavours of Kogut-Susskind
fermions have been obtained
by the HEMCGC Collaboration~\cite{Bitar:1991wr}
at the somewhat more chiral
values~\cite{Bitar:1994rk}, $m_{\pi}/m_{\rho}=0.658(5)$ and
$m_{\pi}/m_{\rho}=0.434(4)$ on lattices with similar
resolution as in the present study, $r_0\approx 5a$.
Their lattice extents are~\cite{Heller:1994rz},
$L_{\sigma}a=3.08(12)\,r_0$ and
$L_{\sigma}=2.50(5)\,r_0$, respectively, and, despite
the rather small volume in the latter case and the
extremely light quarks in the former case, the
results~\cite{Bitar:1991wr,Heller:1994rz},
$m_{0^{++}}=4.47(43)\,r_0^{-1}$ and $m_{0^{++}}=4.17(30)\,r_0^{-1}$,
are compatible with ours.

Recently, the UKQCD Collaboration has reported first results for the
scalar glueball, obtained with the Sheikholeshlami-Wohlert action
on lattices with the linear dimension,
$L_{\sigma}a\approx 3.4\,r_0$,
at chiralities within the range explored in our study,
as indicated by the ratios~\cite{Michael:1999rs},
$m_{\pi}/m_{\rho}\approx 0.71$ and $m_{\pi}/m_{\rho}\approx 0.67$.
However, their lattices are quite coarse: $r_0\approx 3.44\,a$ and
$r_0\approx 3.65\,a$, respectively.
The values they state are remarkably small, namely
$m_{0^{++}}=1.6(2)\,r_0^{-1}$ and $m_{0^{++}}=1.8(4)\,r_0^{-1}$,
i.e.\ they come out to be by a factor two smaller
than our $16^3$ lattice results,
indicating larger finite lattice spacing
effects than seen in quenched simulations~\cite{Bali:2000gf,Michael:1999rs}.
This issue certainly deserves further study.

At sufficiently light sea quark masses we expect the scalar glueball
to mix with either flavour singlet
mesonic states and eventually $\pi\pi$
molecules or to decay into $\pi$ pairs. Such scenarios should be studied,
in un-quenched as well as quenched simulations~\cite{Lee:2000kv,Alford:2000mm}.
In view of the mixing and decay channels that open up once sea quarks are
switched on, it is certainly also worthwhile to investigate glueballs
with other quantum numbers than those that we have studied so far.
In particular the pseudoscalar
should be an interesting candidate,
given the large mass of the $\eta$ meson.
In view of fragmentation models as well as achieving
an understanding of the
decay rates of $\Upsilon(4S)$ and $\Upsilon(5S)$ states into pairs of
$B\bar{B}$ mesons, a determination of the string breaking distance and
the associated mixing matrix elements is certainly equally
exciting. Work along this line is in progress~\cite{Schilling:1999mv,inprep}.

\acknowledgements
This work was supported by DFG grants
Schi 257/1-4, 257/3-2 and 257/3-3 and the DFG Graduiertenkolleg
``Feldtheoretische und Numerische Methoden in der Statistischen und
Elementarteilchenphysik''.
G.B.\ acknowledges support from DFG grants
Ba 1564/3-1, 1564/3-2 and 1564/3-3 as well as
EU grant HPMF-CT-1999-00353.
G.B.\ thanks S.\ Collins for her most useful comments when proof reading
an earlier version of the manuscript.
We thank G.\ Ritzenh\"ofer, P.\ Lacock and S.\ G\"usken
for their contributions
at an earlier stage of this project.
The SESAM HMC productions were run on the APE100 computer at IfH Zeuthen
and on
the Quadrics machine provided by the DFG to the Schwerpunkt ``Dynamische
Fermionen'' operated by the Universit\"at Bielefeld.
We thank V.\ Gimenez, L.\ Giusti, G.\ Martinelli,
and F.\ Rapuano for providing the T$\chi$L configurations which were
mainly produced on an APE100 tower of INFN. We are grateful
to A.\ Mathis for granting considerable computer time at ENEA,
Casaccia.
The analysis was performed on the Cray T90 and J90 systems of the ZAM 
at Forschungszentrum J\"ulich as
well as on workstations of the John von Neumann Institut f\"ur Computing.
We thank the support teams of these institutions for their help.

\end{document}